\documentclass[fleqn,twoside]{article}
\usepackage[headings]{espcrc2}

\readRCS
$Id: espcrc2.tex,v 1.2 2004/02/24 11:22:11 spepping Exp $
\usepackage{graphicx}
\usepackage[figuresright]{rotating}

\newcommand{\bp}{\mbox{\boldmath $p$}}
\newcommand{\bq}{\mbox{\boldmath $q$}}
\newcommand{\br}{\mbox{\boldmath $r$}}

\newcommand{\bDelta}{\mbox{\boldmath $\Delta$}}
\newcommand{\bkappa}{\mbox{\boldmath $\kappa$}}

\newcommand{\bb}{\mbox{\boldmath $b$}}

\newcommand{\AmS}{{\protect\the\textfont2
  A\kern-.1667em\lower.5ex\hbox{M}\kern-.125emS}}

\title{
Unintegrated gluon distribution of a nucleus 
and photon-jet correlations
}

\author{Wolfgang Sch\"afer \address[IFJ]
{
	Institute of Nuclear Physics PAN, 
        ul. Radzikowskiego 152, 31-342 Krak\'ow, Poland
}%
        \thanks{supported in part by the Polish Ministry of Science
and Higher Education under contract 1916/B/H03/2008/34.}
}

\runtitle{Unintegrated gluon distribution of a nucleus}
\runauthor{W. Sch\"afer}

\begin{document}

\begin{abstract}
An impact parameter dependent unintegrated gluon distribution
is constructed as a solution of a nonlinear evolution equation with
realistic Glauber--Gribov input. Photon--jet correlations in $pA$
collisions in the proton fragmentation region are 
proposed as a direct probe of the nuclear unintegrated glue.
\vspace{1pc}
\end{abstract}

\maketitle

\section{Dipole scattering amplitude vs. unintegrated glue}

At high energies, where the relevant transverse momenta
of partons fulfill $\Lambda_{QCD} \ll p_\perp \ll \sqrt{s}$,
we should  take them explicitly into account. The most appropriate
description is one in terms of unintegrated parton (dominantly gluon)
distributions. The latter are related to observables by the 
so-called $k_\perp$-factorization. If we are in a regime where
multiple scattering/saturation effects are large, such as
in the case of a nuclear target, 
the linear $k_\perp$ factorization breaks down and 
observables are in general nonlinear functionals the 
unintegrated glue \cite{Nonlinear,Nonuniversality}.
The nuclear unintegrated glue $\phi(\bb,x,\bp)$ is defined
in terms of the forward scattering amplitude of a $q \bar q$
dipole $\br$ at impact parameter $\bb$, $\Gamma(\bb,x,\br)$:
\begin{eqnarray}
\int\! \! {d^2 \br \over (2\pi)^2} \Gamma(\bb,x,\br)
e^{-i\bp\br}  &=&  (1-w_0(\bb,x)) \delta^{(2)}(\bp) 
\nonumber \\
&-& \phi(\bb,x,\bp) \, ,
\end{eqnarray}
where $w_0(\bb,x)$ has the meaning of Bjorken's gap survival probability.
To construct the nuclear unintegrated glue, one starts at 
moderately small $x \sim x_A \sim 0.01$, where only the $q \bar q$
state is coherent over the whole nuclear size. Then the 
dipole amplitude for the nuclear target takes the simple
Glauber--Gribov form: 
\begin{eqnarray}
\Gamma(\bb,x_A,\br) = 1 - \exp[-\sigma(x_A,\br) T_A(\bb)/2] 
\, 
\end{eqnarray}
where $T_A(\bb)$ is the nuclear matter density. The nuclear 
unintegrated glue can then be expressed as an expansion over
multiple convolutions of the free--nucleon unintegrated glue $f(x,\bp)$
\cite{NSS}:
\begin{eqnarray}
\phi(\bb,x_A,\bp) = \sum w_j(\bb,x_A) f^{(j)}(x_A,\bp) 
\label{expansion}
\end{eqnarray}
Here 
\begin{eqnarray}
 f^{(j)} (x_A,\bp)\! = \! \int \! 
\big[ \prod^j d^2 \bkappa_i \! f(x_A,\bkappa_i) \big] 
\delta^{(2)}( \bp - \sum \bkappa_i) 
\nonumber  
\end{eqnarray}
is the collective glue of $j$ overlapping nucleons, and
\begin{eqnarray}
w_j(\bb,x_A) =
{\nu_A^j(\bb,x_A) \over j! } \exp[-\nu_A(\bb,x_A)] \, ,
\nonumber
\end{eqnarray}
is the probability for $j$ nucleons to contribute.
It depends on the effective opacity
\begin{eqnarray}
\nu_A(\bb,x_A) = {1 \over 2}\alpha_S(q^2)\, \sigma_0(x_A) T_A(\bb)
\, ,
\end{eqnarray}
where $\sigma_0(x) = \int d^2\bp f(x,\bp)$ is a 
nonperturbative parameter, the cross section 
of a large color dipole.
Interestingly, the expansion (\ref{expansion}) also has a simple
unitarity cut interpretation: namely $\phi(\bb,x,\bp)$ is 
proportional to the quasielastic quark--nucleus cross section,
and the $j$-th term in eq.(\ref{expansion}) is the contribution
from $j$ cut Pomerons \cite{Cutting_Rules}.

\section{Nonlinear evolution of the unintegrated glue}
\begin{figure}[htb]
\includegraphics[width=.5\textwidth]{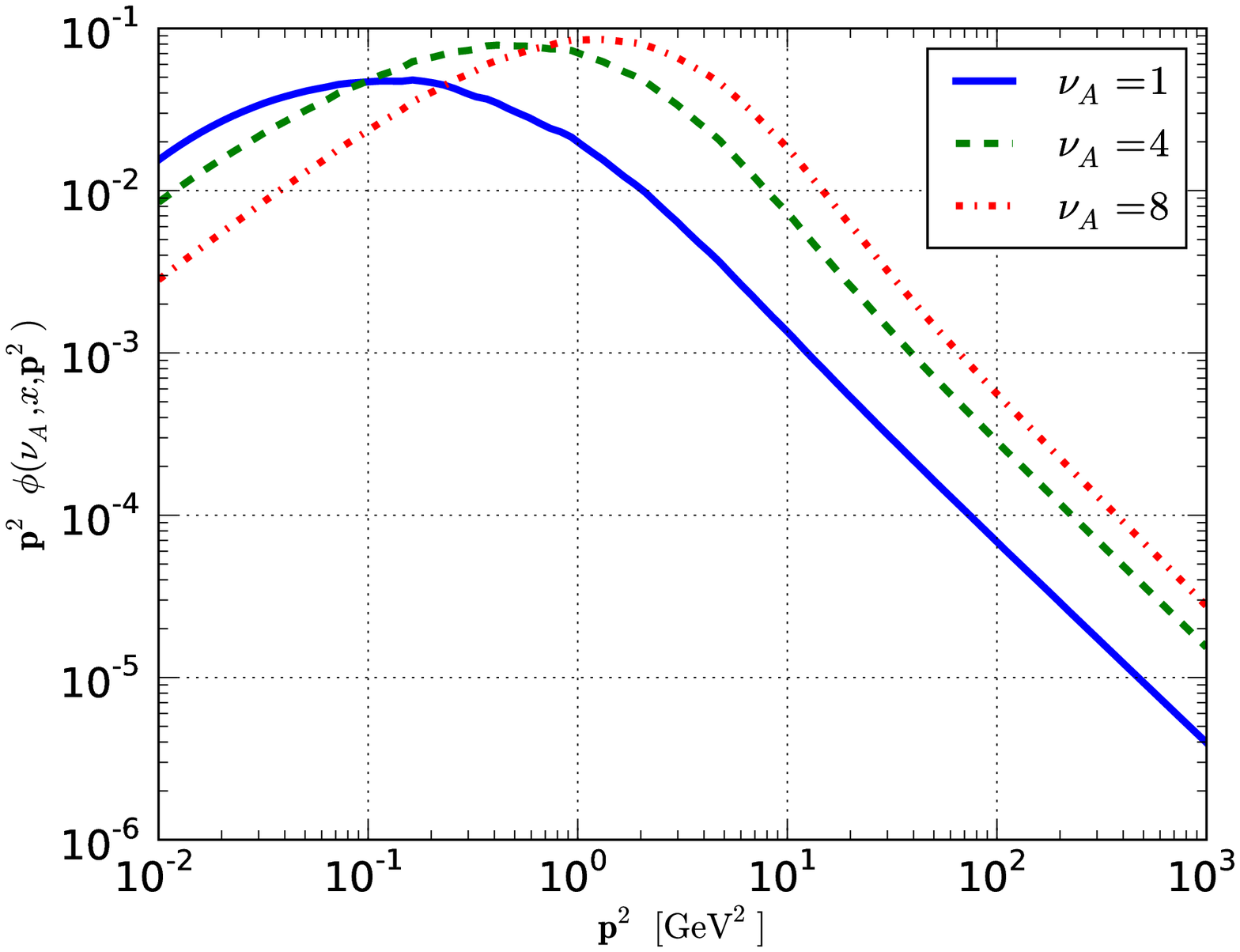}
\includegraphics[width=.5\textwidth]{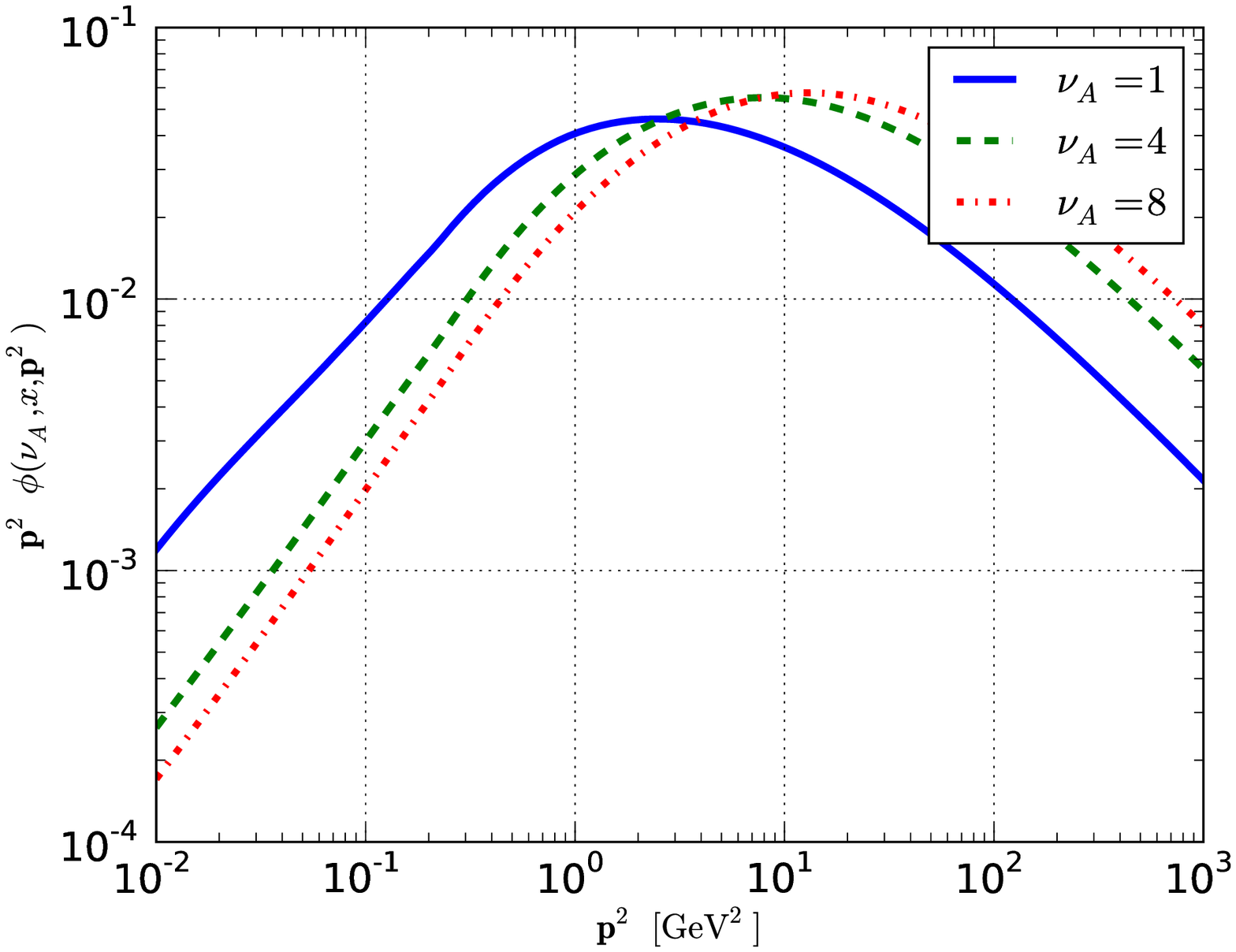}
\vspace{-0.7cm}
\caption{Unintegrated gluon distribution $\bp^2 \phi(\nu_A,x,\bp)$ for 
different opacities $\nu_A$. Top: $x = 0.01$, lower: 
after evolution at $x=10^{-6}$.}
\label{fig1}
\end{figure}
At smaller values of $x$ the multigluon-Fock 
states $q\bar q g, q \bar q g_1g_2 \dots
q\bar q g_1 \dots g_n$ with strongly ordered light-cone 
momentum fractions $z_n \ll z_{n-1} \dots \ll z_1 \ll 1$ 
must be accounted for.
While for the free nucleon target the effect of all multigluon
Fock states is summed up by the linear BFKL--equation, no such 
simple procedure exists for the nuclear target. Simply iterating
the first step of small--$x$ evolution, one would arrive at the
Balitsky--Kovchegov (BK) \cite{BK} equation. In terms of $\phi$ it can be 
put in the form  
\begin{eqnarray}
{\partial \phi(\nu_A,x,\bp) \over \partial \log(1/x)}
&=& {\cal{K}}_{BFKL} \otimes \phi (\nu_A,x,\bp) 
\nonumber \\
&+& {\cal{Q}}[\phi](\nu_A,x,\bp) \, , 
\label{BK}
\end{eqnarray}
with a linear, BFKL, piece, and a quadratic ``gluon fusion''
term ${\cal{Q}}[\phi]$. The latter has a ``real'' as well as
a ``virtual'' piece $\propto \phi(\nu_A,x,\bp)$. The numerical 
treatment is much simplified by the observation, that the real
piece can in fact be written as a square:
\begin{eqnarray}
&&
\int \! d^2\bq d^2\bkappa \phi(\nu_A,x,\bq) 
\Big[ K(\bp+\bkappa,\bp+\bq) 
\nonumber \\
&&- K(\bp,\bkappa+\bp) 
- K(\bp,\bq+\bp)\Big]  
\phi(\nu_A,x,\bkappa) 
\nonumber 
\end{eqnarray}
\begin{eqnarray}
\propto \Big| \int d^2\bkappa \, \phi(\nu_A,x,\bkappa) 
\Big[ {\bp \over \bp^2 + \mu_G^2} - 
{\bp + \bkappa \over (\bp + \bkappa)^2 + \mu_G^2} \Big]
\Big |^2, \nonumber 
\end{eqnarray}
and can be viewed as a diffractive cut of a triple 
Pomeron contribution.
At large $\bp^2$, above the saturation scale, we obtain for the  
fusion term  
\begin{eqnarray}
{\cal{Q}[\phi ]} (\nu_A,x,\bp) \propto -{1 \over \bp^2}
\Big |\int_{\bp^2} {d^2\bkappa \over \bkappa^2} 
\phi(\nu_A,x,\bkappa^2) \Big|^2 \nonumber \\
-  \phi(\nu_A,x,\bp^2) 
\int_{\bp^2} {d^2\bkappa \over \bkappa^2} \int_{\bkappa^2} 
d^2{\bq} \phi(\nu_A,x,\bq^2) \, . \nonumber  
\end{eqnarray}
Notice that it is pure higher twist, and involves only
the ``anticollinear'' integration domain $\bkappa^2 > \bp^2$.
In particular it cannot be written in terms of the 
square of the integrated gluon distribution.

In the practical solution of the evolution equation eq.(\ref{BK})
one integrates over all transverse momenta, from soft to hard, 
and some regularization in the infrared domain is inevitable. 
For example, we freeze the running coupling $\alpha_S$ at small
momenta. It is also necessary to introduce a finite gluon correlation
radius $\mu_G^{-1}$, 
to remove unphysical long-range gluon exchange contributions.
Notice that this can be done without upsetting the 
gauge cancellations in the kernel, as it would be the case e.g.
for a sharp momentum cutoff.

Results from a numerical solution are shown in fig.\ref{fig1}.
The boundary condition at $x_A = 0.01$ was constructed in terms
of a free--nucleon glue fitted to HERA data.
We used $\mu_G^2 = 0.5 \, \mathrm{GeV}^2$. We plot $\bp^2 \phi(\nu_A,x,\bp) 
\propto \partial G / \partial \bp^2 $, which is propotional to the
phase space density of gluons. Instead of impact parameter 
$\bb$ we changed variables to the effective opacity $\nu_A$. 
Larger $\nu_A$ corresponds to more central, and smaller $\nu_A$ 
to more peripheral collisions.

The position of the maximum in fig.\ref{fig1} is 
a good definition of the saturation scale. 
We observe, that it is a function of the opacity as well
as of $x$. It increases for more central 
collisions (where absorption is stronger) as well 
as for smaller $x$. Notice that the evolution of the saturation
scale with $\nu_A$ is a result of the dynamics, and
does not involve an assumption its impact--parameter 
dependence. For the boundary condition, the saturation 
scale is in the soft (for small $\nu_A$) to semihard (for
large $\nu_A$) regime. We need to go to very small 
$x \sim 10^{-6}$ to finally leave the soft region behind.
Notice however that for most observables an average over
$\nu_A$ is implied, and the contribution from the impulse
approximation (or lowest order in $j$) is never really small
(see e.g. \cite{Nonlinear,QuarkGluonDijet}).

\section{Photon-Jet correlations}
\begin{figure}[htb]
\includegraphics[width=.5\textwidth]{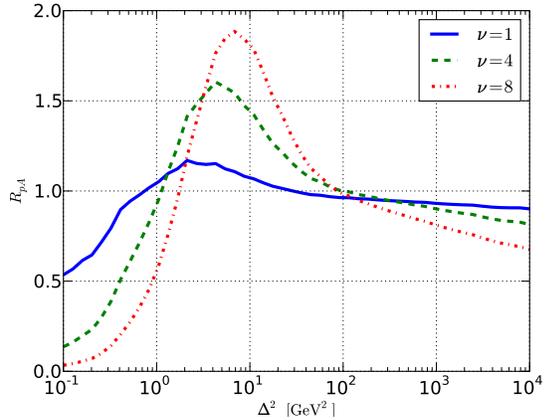}
\includegraphics[width=.5\textwidth]{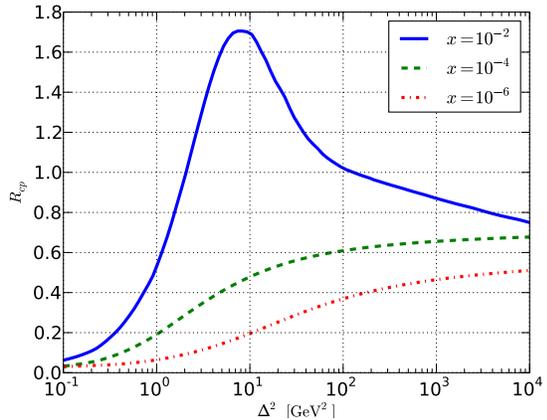}
\caption{Top:$R_{pA}$ at $x=0.01$, Lower:
$R_{cp}$ for $\nu_> = 8, \nu_<=1$ for different $x$.}
\label{fig2}
\end{figure}
It would now be helpful to find an observable which
allows to map the $\bp$--dependence of the unintegrated
glue, and to determine the saturation scale as directly 
as possible.
We have no space here to explain here in detail the 
formalism of nonlinear
$k_\perp$--factorization, which principal feature 
is that cross sections of hard processes in a nuclear 
environment are highly nonlinear functionals 
(quadratures) of the collective nuclear unintegrated glue 
\cite{Nonlinear,SingleJet,QuarkGluonDijet,GluonGluonDijet}.
The emerging nonlinear $k_\perp$-factorization formulas for the hard dijet
spectra we shown to fall into several universality classes depending on 
color properties of the underlying pQCD subprocess 
\cite{Nonuniversality,QuarkGluonDijet,GluonGluonDijet}. 
These classes differ in the character of incoherent initial state and
final state interactions and coherent distortions of two-parton Fock 
states of the incident parton. 
Of special interest are reactions in which both the direct photon and 
the accompanying (balancing, recoiling) quark jet are observed. We shall
refer to such final states as dijets. The 
production of such $q\gamma$ dijets can be viewed, in the nucleus rest frame, 
as an an excitation of the $q\gamma$ Fock state of the incident quark, 
$q^*\to q\gamma$. As with all dijet observables 
\cite{Nonlinear,Nonuniversality,QuarkGluonDijet,GluonGluonDijet}, the
dijet spectrum is obtained in terms of multiparton S-matrices of four
($q \gamma \bar q \gamma$)--,
three($q \bar q \gamma$)-- and two($q \bar q$)--parton states. 
In distinction to purely QCD--processes, however there is a great
simplification due to the fact that the $\gamma$ does not interact
through multigluon exchanges.
As a consequence, the problem {\it{abelianizes}}, and all multiparton
$S$--matrices can be reduced to the ones of $q \bar q$--states.
The emerging dijet spectrum is then a {\it{linear}} functional of the 
unintegrated glue. 

On the free nucleon target, we obtain for the parton--level 
cross section:
\begin{eqnarray}
{2 (2 \pi)^2 d\sigma_N( q \to q \gamma) \over dz d^2\bp d^2\bDelta}
= f(x,\bDelta) \,  P_{\gamma q}(z) 
\nonumber \\
\times K(\bp,\bp - z\bDelta) \, ,
\label{free_nucleon}
\end{eqnarray}
with
\begin{eqnarray}
K(\bp_1,\bp_2)
= \Big| {\bp_1 \over \bp_1^2 + \varepsilon^2}-
  {\bp_2 \over \bp_2^2 + \varepsilon^2} \Big|^2 \, ,
\end{eqnarray}
and the splitting function
\begin{eqnarray}
P_{\gamma q}(z) = 2e_Q^2\alpha_{em}\cdot
{1+(1-z)^2 \over z} \, .  
\end{eqnarray}
Here $z,\bp$ are the photon's light--cone  momentum fraction and
transverse momentum, $\varepsilon^2 = z m_q^2$. 
The decorrelation momentum $\bDelta = \bp + \bp_q =$
measures the deviation of the dijet sustem from the back--to--back situation.
It is worth to point out that the spectrum (\ref{free_nucleon}) is exact 
over the full phasespace of the $q\gamma$ system. In particular
notice the collinear pole at $\bp = z \bDelta$ from the final state 
photon emission of the scattered quark. It corresponds to the ``monojet''
configuration, where the $q\gamma$ system recoils against a jet 
at a large distance in rapidity.

For the nuclear target we obtain:
\begin{eqnarray}
&&{(2 \pi)^2 d\sigma_A( q \to q \gamma) 
\over dz d^2\bp d^2\bDelta d^2\bb}
=
\nonumber \\
&&\Big[ \phi(\nu_A,x,\bDelta) + w_0(\nu_A) \, \delta^{(2)}(\bDelta) 
\Big] 
\nonumber \\
&& \times P_{\gamma q}(z) \,  K(\bp, \bp - z\bDelta) \, .
\end{eqnarray}
A main result of this work is that the decorrelation 
momentum distribution maps out the nuclear unintegrated
glue. 
Notice that a potential diffractive contribution $\propto w_0(\nu_A)$,
which would violate the linear $k_\perp$--factorization, vanishes on a heavy
nucleus, where the momentum transfer distribution is $\delta$-function--like.
For a recent discussion of $\gamma$--particle correlations in the
framework of the Color--Glass--Condensate model, 
see \cite{JalilianMarian}.

Nuclear effects are conveniently characterized by the ratio (here we stay
at the parton--level throughout)
\begin{eqnarray}
R_{pA}(\nu_A,\bp,\bDelta) = {d\sigma_A \over T_A(\bb) d\sigma_N}
= {\phi(\nu_A,x,\bDelta) \over \nu_A \, f(x,\bDelta)} \, .
\nonumber 
\end{eqnarray}
A similar ratio is the central--to--peripheral ratio, which involves
only nuclear quantities:
\begin{eqnarray}
R_{cp}(\nu_>,\nu_<,\bp,\bDelta) = 
{\nu_< \,\phi(\nu_>,x,\bDelta) \over \nu_> \, \phi(\nu_<,x,\bDelta)}
\nonumber
\end{eqnarray}
Interestingly, both these ratios do not depend on the photon's 
transverse momentum $\bp$.
We show the ratio $R_{pA}$ at $x=x_A$ for different opacities 
in the top panel of fig.\ref{fig2}. We observe a shadowing at small values
of $\bDelta$ and a Cronin--type peak which position reflects the 
$\nu_A$-dependent saturation scale.
In the lower panel we show $R_{cp}$ and its evolution with $x$. While
it displays the same Cronin--peak as $R_{pA}$ for $x=x_A$, the latter
is entirely quenched at small $x$.

\section{Conclusions}
We presented a nuclear unintegrated glue from the solution of a 
nonlinear small--$x$ evolution equation. Photon--jet correlations 
in the proton fragmentation region of $pA$ collisions
have been suggested as a direct probe of the nuclear unintegrated glue.
They may present an opportunity to measure the saturation scale at the LHC.

\section{Acknowledgements}
It is a pleasure to thank Kolya Nikolaev for collaboration, 
especially on Sec.3, which is based on our unpublished work 
from 2005.


\begin{thebibliography}{9}

\bibitem{Nonlinear}
  N.~N.~Nikolaev, W.~Sch\"afer, B.~G.~Zakharov and V.~R.~Zoller,
  J.\ Exp.\ Theor.\ Phys.\  {\bf 97} (2003) 441.

\bibitem{Nonuniversality}
  N.~N.~Nikolaev, W.~Sch\"afer and B.~G.~Zakharov,
  Phys.\ Rev.\ Lett.\  {\bf 95} (2005) 221803.

\bibitem{NSS}
  N.~N.~Nikolaev, W.~Sch\"afer and G.~Schwiete,
  Phys.\ Rev.\  D {\bf 63} (2001) 014020.

\bibitem{Cutting_Rules}
  N.~N.~Nikolaev and W.~Sch\"afer,
  Phys.\ Rev.\  D {\bf 74} (2006) 074021.

\bibitem{BK}
  I.~Balitsky,
  Nucl.\ Phys.\  B {\bf 463} (1996) 99;
  Y.~V.~Kovchegov,
  Phys.\ Rev.\  D {\bf 60} (1999) 034008.

\bibitem{QuarkGluonDijet}
  N.~N.~Nikolaev, W.~Sch\"afer, B.~G.~Zakharov and V.~R.~Zoller,
  Phys.\ Rev.\  D {\bf 72} (2005) 034033.

\bibitem{SingleJet}
  N.~N.~Nikolaev and W.~Sch\"afer,
  Phys.\ Rev.\  D {\bf 71} (2005) 014023.

\bibitem{GluonGluonDijet}
  N.~N.~Nikolaev, W.~Sch\"afer and B.~G.~Zakharov,
  Phys.\ Rev.\  D {\bf 72} (2005) 114018.

\bibitem{JalilianMarian}
  J.~Jalilian-Marian,
  Eur.\ Phys.\ J.\  C {\bf 61} (2009) 789.


\end{thebibliography}
\end{document}